# IMS-Based Mobile Learning System

M. Rizwan Jameel Qureshi

Faculty of Computing and Information Technology, King Abdulaziz University, Jeddah, Saudi Arabia
anriz@hotmail.com

**Abstract:** Electronic (E) learning management system is not a novel idea in the educational domain. Learning management systems are used to deal with academic activities such as course syllabi, time table scheduling, assessments and project discussion forums. Almost, all the top universities of world are using general purpose/customized solutions to manage learning management systems like SAP, Oracle, Moodle and Blackboard. The aim of this paper i.e., Mobile (M) Learning System (MLS) is not to substitute the traditional web based E learning applications but to enhance it by amalgamating both web and mobile technologies. This idea justifies the proposal of M learning system to use some of the services of E learning system from mobiles. MLS will use state-of-the-art IP Multimedia Sub System technology. The emphasis in this research will be on the technical implementation of the Session Initiation Protocol (SIP) using IP Multimedia Subsystem (IMS) to develop an MLS not only for the students of the King Abdulaziz University but it will be beneficial for the students of other universities at Kingdom of Saudi Arabia. A customized CBD is proposed as per the nature of MLS project. MLS case study is used as a research design to validate the customized CBD model. Multi-tier applications architecture (client, web, and business) will be adopted during the development of MLS case study. An MLS will be developed and tested using IMS platform to check its practicality for the students of King Abdulaziz University. It is anticipated that the proposed system will significantly facilitate to both the students and teachers of KAU during their off campus activities.
[Qureshi MRJ. **IMS-Based Mobile Learning System**. *Life Sci J* 2013;10(4):2121-2126]. (ISSN:1097-8135). http://www.lifesciencesite.com. 282

**Keywords:** M-learning, E-learning; IMS platform; SIP protocol

## 1. Introduction

There is a significant shift in the academics during the last few years from traditional teaching to mobile learning environment with the advancement of mobile wireless technologies. Several universities incorporated the mobile technology into their teaching and learning environment by recognizing the potentials of mobile technology as an effective medium. There are several research works published by Chihab (2003), Bollen et al. (2004), and Bowman and Bowman (1998) on the effectiveness of Mobile (M) learning systems with the electronic (E) learning systems. However, little research has been done on effect of the use of ubiquitous devices for teaching and learning. The objective of this research is not to replace traditional E learning system with M learning. But, M learning can go together with E learning to add value to the learning and administrative activities.

Mobile Learning System (MLS) is developed during this research to solve the problem taken up in this paper. MLS uses the IP based Multimedia subsystem (IMS) platform to deliver the services. The main aim of IMS is to facilitate the Internet services to the user anywhere and anytime through the mobile phone technology. IMS is designed with building blocks enabling telecommunication operators to deliver new services in a more flexible way. IMS networks bring a seamless delivery service model to the server providers such as messaging, video, data apart from the type of the network to achieve better quality of service (QOS). IMS network architecture has three main layers that are transport, control and service. These layers make it easy to standardize the interfaces and interconnect the systems [3]. King Abdulaziz University is using many web applications such as On Demand University system (ODUS) Plus, Moodle, ANJEZ and CENTRA. Users (students) are facing many problems with the usage of these applications such as students don't know how to add & drop the courses, can't find the buildings of lectures, and can't remember the events like exams, home works and lectures. Faculty also faced several similar difficulties regarding the interaction with the students through the conventional web based systems. The objective of this paper to propose an MLS as a case study to investigate how far the concept of mobile learning is accepted by the students and faculty members in King Abdulaziz University and result are reported in this paper.

Section 2 covers the related work. The problem selected in this paper is covered in section 3. Section 4 describes the details of customized component based development (CBD) model to develop MLS. The validation of MLS is illustrated in section 5.

## 2. Related Work

The literature review is composed of two parts. First part is used to identify the existing M learning systems to support this research. Existing models are studies in the second part to identify the problems in the CBD domain to propose a customized CBD model for this research.

### 2.1 Existing M Learning Systems

Colorado State and Sussex universities are using M-learning systems. Colorado State University (CSU) is using an iPhone application to provide mobile access to its users for updated information of courses, faculty, athletics, calendar of events and videos. Whereas, Sussex University has introduced M-learning application for Android, iPhone and Blackberry platforms. Both applications provide a variety of services such as search a faculty member's telephone number or email and upcoming events to appraise courses in the catalog. Building locator function is used to search buildings of campus to locate a room. Mobile applications have a directory service to search and save contacts of faculty and staff on mobile to email or call. Another service is providing the latest information about educational, cultural, and social events at campus. On campus athletics news with detailed schedule and updated scores are also available in mobile application. Library service is also accessible through mobile applications to read electronic books.

Alier et al. (2007) integrated mobile learning application with learning management systems. Alier and his colleagues are of the view that mobile learning is not a replacement of traditional web based electronic learning. Integration techniques and interoperability standards are also discussed. A case study is conducted to develop Moodbile application using IMS (Alier et al., 2007). The main limitation of Alier et al. (2007) work is unavailability of results that what outcomes are achieved, after the implementation of Moodbile application, and what are the responses of users.

### 2.2 Existing CBD Models

A software-cycle model is proposed for reuse and reengineering (John and Victor, 1991) suggests five stages to reuse a component.

- Analysis of existing programs to sort components to be reused.
- Reengineering to eliminate domain specific troubles.
- Saving reusable components in the repository.
- Construction of independent status components with a reuse approach to store in a repository.
- Reuse components to develop new programs.

The proposed model is not a complete process model for CBD but focus of this research is on reuse activities. Poorly gathered SW requirements could fail a SW project (Hemant et al., 2003). It is because of natural drawback in requirements determination methods. An approach is proposed to construct SW by categorizing components in a knowledge base. Existing components are recognized, chosen and integrated in a newly developed SW by using the knowledge base. Different classification schemes to reuse artifacts have been discussed as well. These are enumerated, keyword, faceted and hypertext. The objective of this paper is to ease requirements gathering using knowledge base but the paper lacks in suggesting a comprehensive process model for CBD.

An incremental method is presented for distributed CBD (de Almeida et al., 2004). It is based on two phases. The first phase composes of gathering requirements of the problem domain and construction of employable components in object-oriented (OO) language. These employable components are stored in a repository. SW Engineers look up MVCASE tool to select the necessary components to develop the SW system in the second phase. The proposed process model is not a transparent model and a use of a specific CASE tool is the requirement of this process model.

A software life cycle model is anticipated to support CBD using OO construction (Luiz et al., 2001). The main phases of model are Domain Engineering, System Analysis, Design and Implementation. The major problem of this model is the selection of reusable components during the design phase. The selection of reusable components should be during the analysis phase. Therefore analyst can estimate the cost, schedule and effort required to develop and integrate the components.

The four stage component-based development process model is very complex for implementation (Hutchinson et al., 2004). The core objective of this paper is to integrate off-the-shelf components with the newly developed components rather than in house development. Repository has not been used here.

CBD process model main phases are Component Analysis, Architectural Design, Component Brokerage, Component Production and Component Integration phases (Ning, 1996). The CBD model is a modification of Waterfall process model with the integration of these phases. Waterfall model is not suitable for commercial applications because of the verification of phase's repetition. It is time and cost consuming process model which is suitable only for research projects.

## 3. Problem Definition

The study area is focused on transition of some of the web services to mobile services for learning purposes. For this purpose, a customized CBD model is proposed. The research question, taken up in this paper, is as follows

"*How to propose a customized CBD model develop IMS-based M learning system to facilitate the students in their off campus and on campus activities*?"

## 4. The Proposed Customized CBD model

The Component Based Development (CBD) approach is used to develop IMS-Based M learning system. The CBD approach is selected because IMS architecture is based on components. The figure 1 depicts the CBD architecture of IMS platform. IMS works on using five layers model that are Application, Service, Control, Transport and Device. Each layer is using components to exchange information to pass from one layer to another. The details of IMS layers are not the scope of this paper. A customized CBD model is proposed to develop MLS. Figure 2 is the diagrammatic representation of the proposed customized CBD model. The details of customized CBD model phases are as follows.

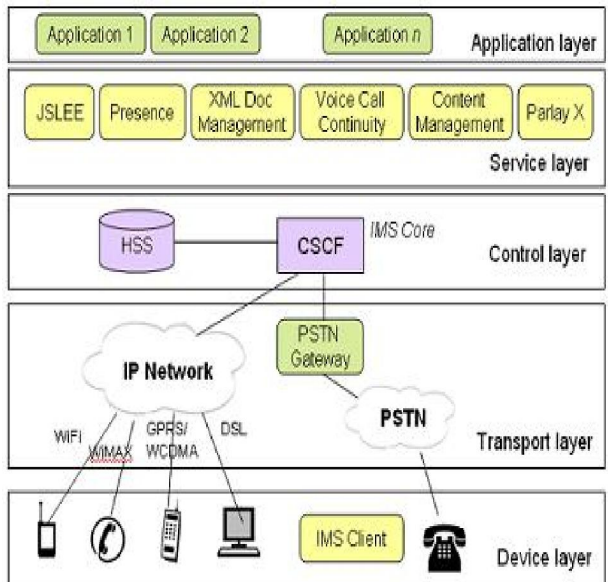

Figure 1. CBD Architecture of IMS Platform (Huang, 2009)

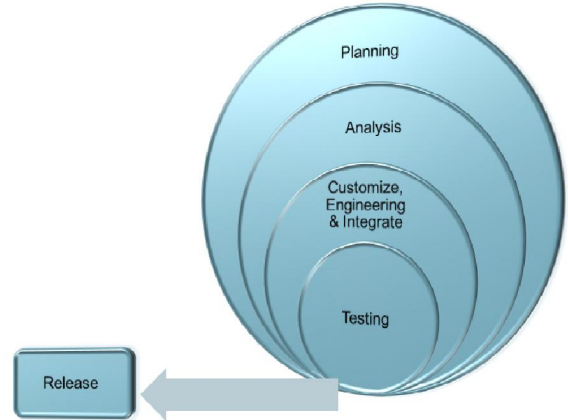

Figure 2. The Proposed Customized CBD Model

**Planning Phase**- Students of King Abdulaziz University (KAU) are communicated at the start of the project to gather basic requirements.

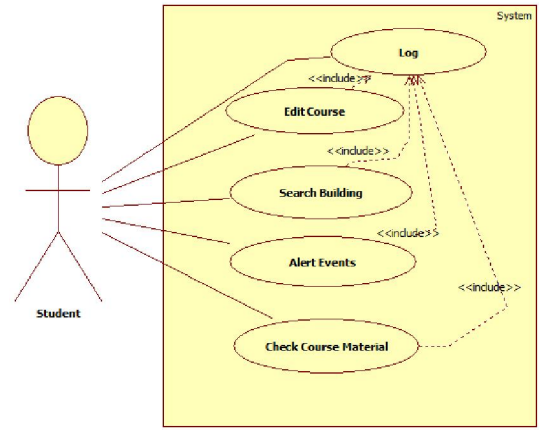

Figure 3. The Use Case Diagram of M Learning System

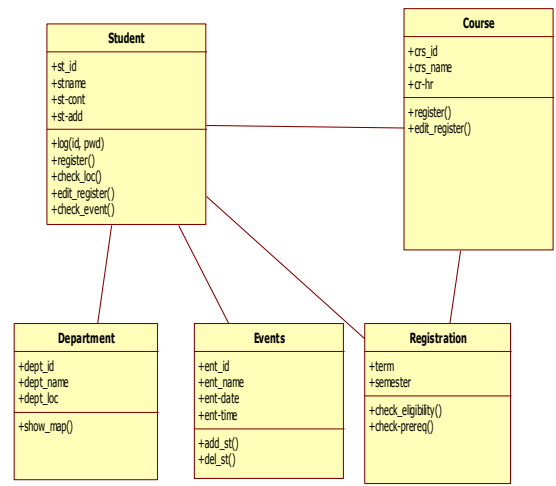

Figure 4. The Class Diagram of M Learning System

Initial use cases are developed at this stage to prepare project specification. Project specification or proposal document is composed of feasibility and risk assessments that are performed to prepare a cost benefits analysis (CBA) sheet. CBA sheet helps the KAU management to decide whether the M learning project is feasible for the KAU or not.

**Analysis phase**- The detailed analysis is started after the top management of KAU approved the proposal of the said system. Functional specification is end product of Analysis phase. The main contents of functional specification are detailed description of system's requirements, functional modeling using the Unified Modeling Language (UML) and data modeling using entity relationship diagram (ERD). In order to achieve detailed requirements a survey is distributed among hundred (100) students of Faculty of Computing & Information Technology (FCIT), KAU. Seventy (70) out of hundred (100) questionnaires are returned. The students are selected randomly (following the random sampling technique) from $1^{st}$ semester to $9^{th}$ semester. Ten students of $9^{th}$ semester voluntarily involved in this project to act as the role of customer. $9^{th}$ semester students were the most senior and they had covered all the courses of Software Engineering track. These ten students, who participated in the MLS project, are written hence onwards in this paper as customer. The requirements are finalized after analyzing the data of questionnaires using the SPSS software and using extensive prototyping to customer. A domain analysis is performed to find a suitable architecture for the application to be developed (Luiz et al., 2001). An architectural model of application is developed that enables a software engineer to evaluates efficiency of design, judge options of design and minimize potential threats coupled with software development (Pressman, 2010). The properties, behavior and relationships among components are identified. Figure 2 and figure 3 show the Use Case and Class diagrams of M learning system. ERD is designed using the MS Visio.

Student will interact with IMS network using his mobile. Logger interface creates session if successfully validated by IMS network to provide access to application. The architecture developed for the MLS is using the CBD approach i.e., each component has its own job. Following are the details about the main components of MLS that are identified during the 'Analysis' phase of customized CBD model.

Network Registration Component is responsible for Network registration to validate the student using IMS platform. The student registration procedure deals with mutual authentication between the user equipment (UE) and the IMS platform and it uses the 3GPP Authentication and Key Agreement (3GPP-AKA) protocol (Zhang, (2005)). The process of authentication is performed by using IMS core infrastructure (IMS enabler) comprising control functions and subscriber database. Administration component defines different users of the M learning system with their designated roles. Interface component publishes graphical user interfaces (GUIs) for the users. 'Search Building' component will get the information from the student about the building number and name to measure the longitude and latitude of the buildings. After that the 'Search Building' component will connect to Global Positioning System (GPS) to measure the location of student and display the location of building. 'Edit Course' component allows the student to access the ODUS Plus database to add and drop the courses using the MLS. 'Check Course' component is responsible to access the course material and assignments. 'Check Event' component is responsible to check social events and academic activities like exam and classes schedule, enrollment dates for next semester and information about the college events.

**'Customize, Engineering and Integrate' Phase- 'Customize'** part of CBD model will be used in future releases of this project because the MLS is developed from the scratch. **Engineering** part is composed of designing and development. Technical specification document is completed during the designing. Technical specification is composed of interface specification, flowchart and database construction.

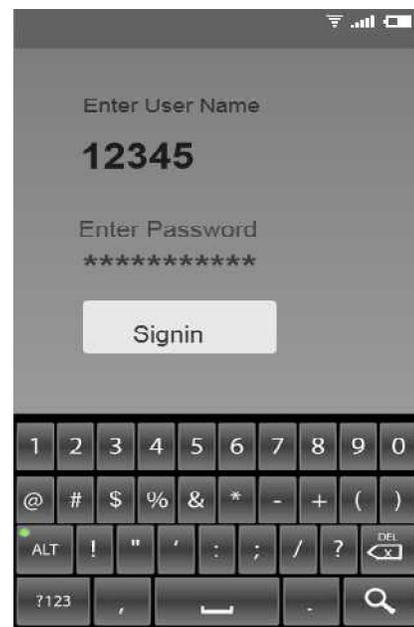

Figure 5. Log Interface of M Learning System

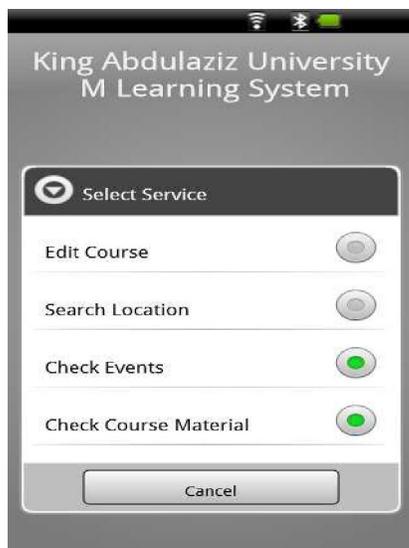

Figure 6. Main Window of M Learning System

Prototyping technique is used to approve the interface and functions. Review of technical specification is also performed based on the feedback of students. Figures 5 and 6 show 'Logon' and 'Main Window' screens of MLS using Android platform. The remaining screens of system are not shown in this paper.

Development is started after completing the technical specification. The required technology for this project is mobile, server machine, IMS, Java JDK, Android SDK and Google Map and application programming interface (API). API is developed to integrate M leaning system running on IMS platform and web based ODUS Plus system. API acts as base to provide web services using the MLS. **Integrate** component/s to system immediately after testing of programmers and quality assurance team. The configuration manager labels each component after the approval of project manager.

**Testing Phase**- MLS is developed using CBD approach. Unit, integration, system and acceptance tests are performed throughout the development of MLS. Customer is called for testing at software company side for alpha testing. Maintenance is performed based on feedback of customer. Beta testing is performed at FCIT KAU site to check the main functionalities of MLS. Maintenance is again performed based on uncontrolled testing and first release is ready to use. The customized CBD model is iterative and MLS is developed in increments.

## 5. Validation of the Customized CBD Model

A case study of MLS, as a research method, is used to validate the customized CBD model. The case study is conducted in the vicinity of software company. A team of seven members, out of forty members, were selected for this project based on the domain experience and knowledge. The development team has medium to high experience ranging from three twelve years. The team composed of one project manager, one technical architect, one configuration manager, three programmers and one quality assurance engineer. The three programmers will act the role of database designers as well. The role of author in this project is of a researcher. The software company is already practicing a Rational Unified Process (RUP) model for existing projects. One week training is arranged for the development team about the customized CBD model main processes and activities.

The results of case study are reported based on the participant and direct observations. The results are reported based on the first four releases of MLS. Table 1 shows the summary of data for the first four releases of case study project. It is shown in Table 1 that the size of first release is 3840 lines of code (LOC) within duration of 3 months. Fault rate per kilo line of code (KLOC) is 4.46. Formal technical review (FTR) technique is used during the testing with a frequency of 3% of the project time. Customer acceptance frequency is 60% for the first release of MSL with a 40% changes in the requirements.

In Table 1, the sizes of next three releases are 4000, 4250 and 4360 LOC. Two months are spent to complete the next three releases. 3.98, 3.40 and 2.76 fault rates per KLOC are reported for the last three releases. The FTR frequency rates are 2%, 1.75% and 1.5% for the $2^{nd}$, $3^{rd}$ and $4^{th}$ releases. The customer acceptance rates are 75%, 77% and 80% for the last three releases. The changes in customer requirements after acceptance tests are 25%, 23% and 20% for the last three releases. The validation of the first four releases of MLS shows the effectiveness of customized CBD model.

Table 1. Summary of data for the first four releases

| Releases | 1 | 2 | 3 | 4 |
|---|---|---|---|---|
| Size in LOC | 3840 | 4000 | 4250 | 4360 |
| Time (months) | 3 | 2 | 2 | 2 |
| Fault/KLOC | 4.46 | 3.98 | 3.40 | 2.76 |
| Formal Technical Review Frequency | 3% | 2% | 1.75% | 1.5% |
| Acceptance Test | 60% | 75% | 77% | 80% |
| Change Ratio on Feedback after Testing | 40% | 25% | 23% | 20% |

## 6. Conclusion

The main objective of this research is two folds by proposing a customized CBD model and developing an MLS to solve the problems of KAU students. The students of KAU are facing many problems using an existing web based solution to add and drop courses, difficulty in finding the buildings and class rooms and facing problems to search the events during their off campus activities. IMS platform is used to provide quality of service using MLS to solve the problems of students. CBD approach is followed to develop MLS because IMS is using layered architecture and each layer is an independent component. A customized CBD model is proposed by addressing the problems of existing CBD models during this research. The results indicate that customized CBD model works effectively to develop the first four releases of MLS. A clear cut improvement is observed from release to release due to improvement in software processes, enhanced learning of team about the project and high involvement of user. The customer acceptance rate shows the significant quality of services using the MLS due to IMS platform. The integration of MLS with the web based system of KAU is achieved to provide some of the web services using the mobile technology to solve the problems of students.

**Acknowledgements:**
This work was funded by the Deanship of Scientific Research (DSR), King Abdulaziz University, Jeddah, under grant No. (12-611-D1432). The author, therefore acknowledge with thanks DSR technical and financial support.

**Corresponding Author:**
Dr. M. Rizwan Jameel Qureshi
Department of Information Technology
Faculty of Computing & Information Technology,
King Abdulaziz University
80221 Jeddah 21589, Saudi Arabia
E-mail: rmuhammd@kau.edu.sa

**References**
1. BenMonssa C. Workers on the move new opportunities through mobile commerce. Proc. of the IADIS International Conference, Lisbon, Portugal, 3-6 June 2003, 2003.
2. Bollen L, Eimler S, Hoppe HU. SMS-based discussions - technology enhanced collaboration for a literature course. Proc. of 2nd IEEE International Workshop on Wireless and Mobile Technologies in Education (WMTE'04), 2004.
3. Bowman RL, Bowman VE. Life on the electronic frontier: the application of technology to group work. Journal for Specialists in Group Work 1998;23(4):428-45.
4. Huang S. Mobile Telemedicine System based on IMS/SIP platform, M.Sc. Thesis, School of Information and Communication Tech., Royal Institute of Technology, Stockholm, Sweden, 2009.
5. Alier M, Casany M, Casado P. A Mobile extension to a web based Moodle virtual classroom. IOS Press 2007;4:1169-76.
6. John WB, Victor RB. The Software-Cycle Model for Reengineering and Reuse. Proc. Of today's accomplishments; tomorrow's expectations USA 1991:267-81.
7. Hemant J, Padmal V, Fatemah MZ. An Assessment Model for Requirements Identification in Component-Based Software Development. The DATA BASE for Advances in Information Systems 2003;34(4):48-63.
8. de Almeida ES, Alvaro A, Lucredio D, do Prado AF, Trevelin LC. Distributed Component-Based Software Development: An Incremental Approach. Proc. of 28th Annual International Conference on Computer Software and Applications (COMPSAC) 2004, Hong Kong, China:4-9.
9. Luiz FC, Miriam AMC, Dahai L. Component-Based Software Process. Proc. 7th Int. Conf. Object-Oriented Information Systems, Calgary, Canada 2001:523-529.
10. Hutchinson J, Kotonya G, Sommerville I, Hall S. A Service Model for Component-Based Development. Proc. 30th EUROMICRO Conf., Rennes- France 2004:162-169.
11. Ning JQ. A Component-Based Software Development Model. Proc. 20th Conf. Computer Software and Applications, Seoul, Korea 1996: 389-394.
12. Pressman RS. Software Engineering. McGraw Hill Press, USA, 2010.
13. Zhang M, Fang Y. Security Analysis and Enhancements of 3GPP Authentication and Key Agreement Protocol. IEEE Transactions on Wireless Communications 2005;4(2):734-42.

7/1/2013